\documentclass[11pt]{article}
\usepackage{moriond,epsfig,amssymb}

\def\be{\begin{equation}}
\def\ee{\end{equation}}
\def\bea{\begin{eqnarray}}
\def\eea{\end{eqnarray}}

\begin{document}
\vspace*{4cm}
\title{EXPLORING THE UNIVERSE WITH THE ANTARES NEUTRINO TELESCOPE}

\author{V. A. KUDRYAVTSEV for the ANTARES Collaboration}

\address{Department of Physics and Astronomy, University of Sheffield,
Sheffield S3 7RH, UK}

\maketitle\abstracts{
The ANTARES Collaboration is currently constructing a large neutrino
telescope in the Mediterranean sea. The telescope will use a 
three-dimensional array of photomultiplier tubes (PMTs) to detect the
Cherenkov light emitted in sea water by neutrino-induced muons.
The array of 12 strings, which hold 900 PMTs,
is planned to be deployed at a depth of about 2500 m near Toulon
(France), 40 km off the coast.
The scientific objective of ANTARES is to detect
high-energy neutrinos, which may be produced at astrophysical
sources -- sites of the cosmic-ray acceleration, such as quasars,
gamma-ray bursters, microquasars and supernova remnants. The
objectives also include the indirect search for WIMPs by looking for 
neutrinos from neutralino annihilations in the centres of 
the Sun, the Earth and the Galaxy.
In 2003 two strings have been successfully deployed and
connected to the electro-optical cable, which transmitted
data to shore station.
}

\vspace{-0.5cm}
\section{Introduction}
High-energy neutrino astronomy opens a new window on the Universe,
complementary to existing gamma-ray observations. Neutrinos can 
be produced in {\it pp} or {\it p$\gamma$} interactions of 
accelerated protons (or heavier nuclei) with matter or photons
via the decay of charged pions (and possibly kaons). Neutrinos 
escape from the source and travel large distances to the Earth 
without being absorbed, scattered or deflected by magnetic fields.
Thus they can deliver information directly from the sites of cosmic-ray 
acceleration, whereas high-energy gammas and protons can be absorbed 
either close to the source or on their way to the Earth, and protons 
are also deflected by magnetic fields. 
High-energy neutrinos can also be produced in the annihilation of 
neutralinos (favourite supersymmetric candidate for non-baryonic dark matter) 
accumulated in the centres of celestial bodies.

Several projects are now underway to construct large-scale neutrino 
detectors underwater or under ice, such as Baikal, AMANDA, NESTOR, ANTARES, 
NEMO, IceCube \cite{halzen}, with Baikal and AMANDA already running
and producing interesting results.

\section{Detector design}

\begin{figure}[htb]
\begin{center}
\includegraphics[width=12.0cm,height=6.0cm]{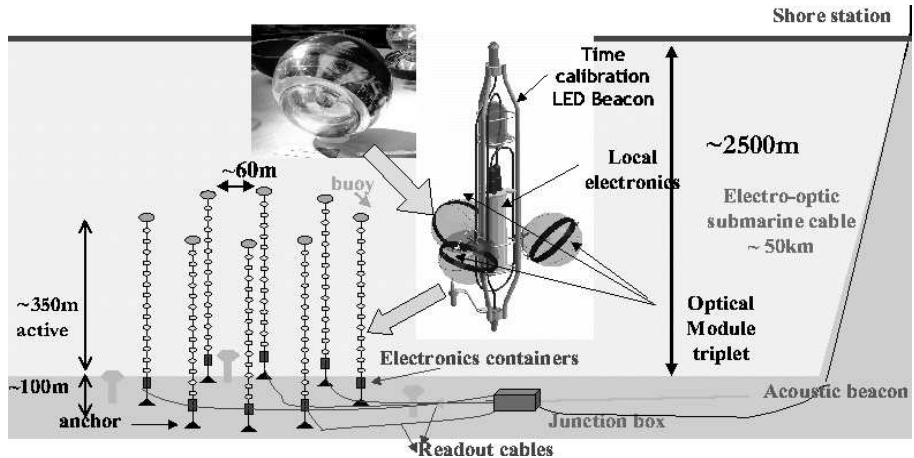}
\caption{Schematic view of the ANTARES detector.}
\label{fig-antares}
\end{center}
\end{figure}

ANTARES ({\bf A}stronomy with a {\bf N}eutrino {\bf T}elescope 
and {\bf A}byss environmental {\bf RES}earch) \cite{proposal} 
will detect the
Cherenkov light emitted by secondary particles produced in neutrino 
interactions in
sea water or the rock below the sea bed. The detector is optimised for
detecting muons from 
charged-current reactions of muon neutrinos, but
will also be sensitive to other neutrino flavours and to neutral-current
reactions. It will consist of 12 lines (or strings) which will be anchored to
the sea bed at distances of about 60-75 m from each other and kept vertical by
buoys. Each string will be equipped with 75 optical modules (OMs) \cite{om}
arranged in triplets (storeys, see Figure~\ref{fig-antares}) subtended by
titanium frames which also support water-tight titanium containers for the
electronic components. Each OM glass sphere houses one 10-inch 
PMT, oriented at $45^\circ$ 
to the downward vertical. The storeys will be spaced at a vertical distance of 
14.5~m and
interconnected with an electro-optical-mechanical cable supplying the electric
power and the control signals, and transferring the data to the bottom 
of the string.
Submarine-deployed electro-optical cables connect the strings to the
junction box (JB) which then sends the data to shore via the main electro-optical
cable. Each string will carry several optical beacons for timing
calibration and acoustic transponders used for position measurements. The
detector will be complemented by an instrumentation line supporting
devices for measurements of environmental parameters and tools used by other
scientific communities (for instance, a seismometer).

The PMT signals are processed
with custom-designed ASIC chips which measure the
arrival time and charge for signals. All signals above an adjustable
threshold (usually corresponding to a fraction of one photoelectron 
pulse) are sent to shore,
where an online filter running on a PC farm selects event candidates and reduces
the data recorded on tape to about 1 MB/s.

\vspace{-0.2cm}
\section{ANTARES performance and sensitivity}

Muons produced by muon neutrinos are finally identified and
reconstructed by offline algorithms. 
Relative and absolute positions of OMs will be measured with an accuracy
of $\sim 5$ cm and $\sim 1$ m, respectively.
From the arrival times of the photons at
the PMTs and the OM positions, the trajectories of muons
will be reconstructed. The resulting angular resolution
estimated from simulations is about $0.2^\circ-0.3^\circ$
for neutrino energies $E_{\nu} > 10$ TeV. At smaller energies 
the angle between the reconstructed muon and the parent
neutrino direction is dominated by the kinematics of the
charged current interaction.
The muon energy, $E_{\mu}$,
is determined from the muon range at small energies and from the Cherenkov
intensity due to radiative energy losses at high energies.
The energy resolution is about $30-40\%$ in log$E_{\mu}$ for $E_{\mu} > 10$ TeV.
The effective area of ANTARES depends on the neutrino 
energy, the efficiency of reconstruction and selection cuts. For well
reconstructed muon tracks (accuracy better than $1^\circ$) 
the effective surface area increases
from about $5 \times 10^{-3}$ km$^{2}$ at $E_{\nu} \sim 0.1$ TeV to more than 
$0.05$ km$^{2}$ at $E_{\nu} > 100$ TeV. The Earth, 
however, becomes opaque to very-high-energy neutrinos.

Atmospheric down-going muons and up-going 
neutrinos constitute the physical background for ANTARES. 
Atmospheric muons, in particular muon bundles, can be erroneously 
reconstructed as up-going muons expected from neutrino interactions.
This background is currently under study, preliminary 
results showing that the large depth and sophisticated reconstruction 
algorithm can help to suppress it sufficiently to 
be sensitive to astrophysical neutrinos. The signal from atmospheric 
neutrinos is indistinguishable from astrophysical neutrinos. This 
background, however, is negligible in a search for point sources of 
high-energy neutrinos, provided the angular resolution of 
$< 1^\circ$ can be achieved in practice. For diffuse neutrino fluxes 
the spectral information becomes crucial with a diffuse spectrum 
being expected to be harder (power index $\gamma \approx -2$) than 
conventional atmospheric neutrino spectrum ($\gamma \approx -3.6$ for 
$E_{\nu} \gtrsim 1$ TeV).

Detailed simulations have been carried out to assess the physics
sensitivity of ANTARES. After 3 years of operation, the ANTARES data will
challenge predicted upper limits for diffuse neutrino fluxes
\cite{diffuse} and will be sensitive to point source intensities
predicted by different models. In Figure~2, the expected ANTARES
sensitivity to muon flux induced by neutrinos from point sources is compared to
upper limits from other experiments (see the review \cite{teresa} for details
and references). Note in particular the complementary sky
coverage of AMANDA and ANTARES. 

The search for neutrinos from gravitational
centres, such as the Sun, the Earth and the Galactic Centre, 
yields sensitivity to WIMP
annihilation and thus complements direct searches for non-baryonic 
dark matter. Figure~3 shows expected ANTARES sensitivity 
to the muon flux from neutralino annihilations 
in the centre of the Sun for the case of `hard' neutrino spectrum
\cite{dm} (assuming $100\%$ annihilations to {\it WW}).

\vspace{-0.5cm}
\begin{figure}[htb]
\begin{center}
\includegraphics[width=6.5cm,height=5.5cm]{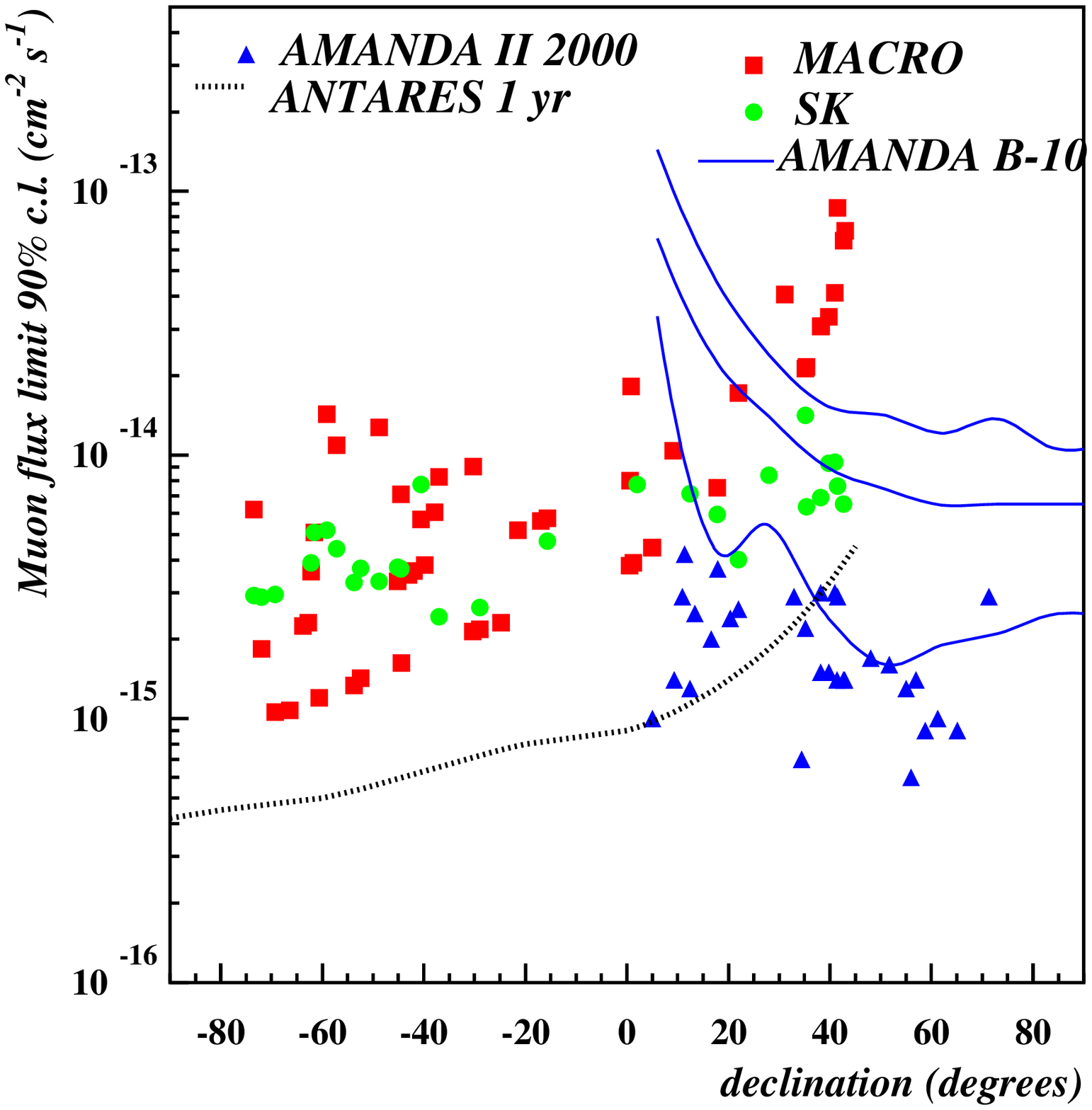}
\label{fig-point}
\hspace{1.0cm}\includegraphics[width=6.5cm,height=5.3cm]{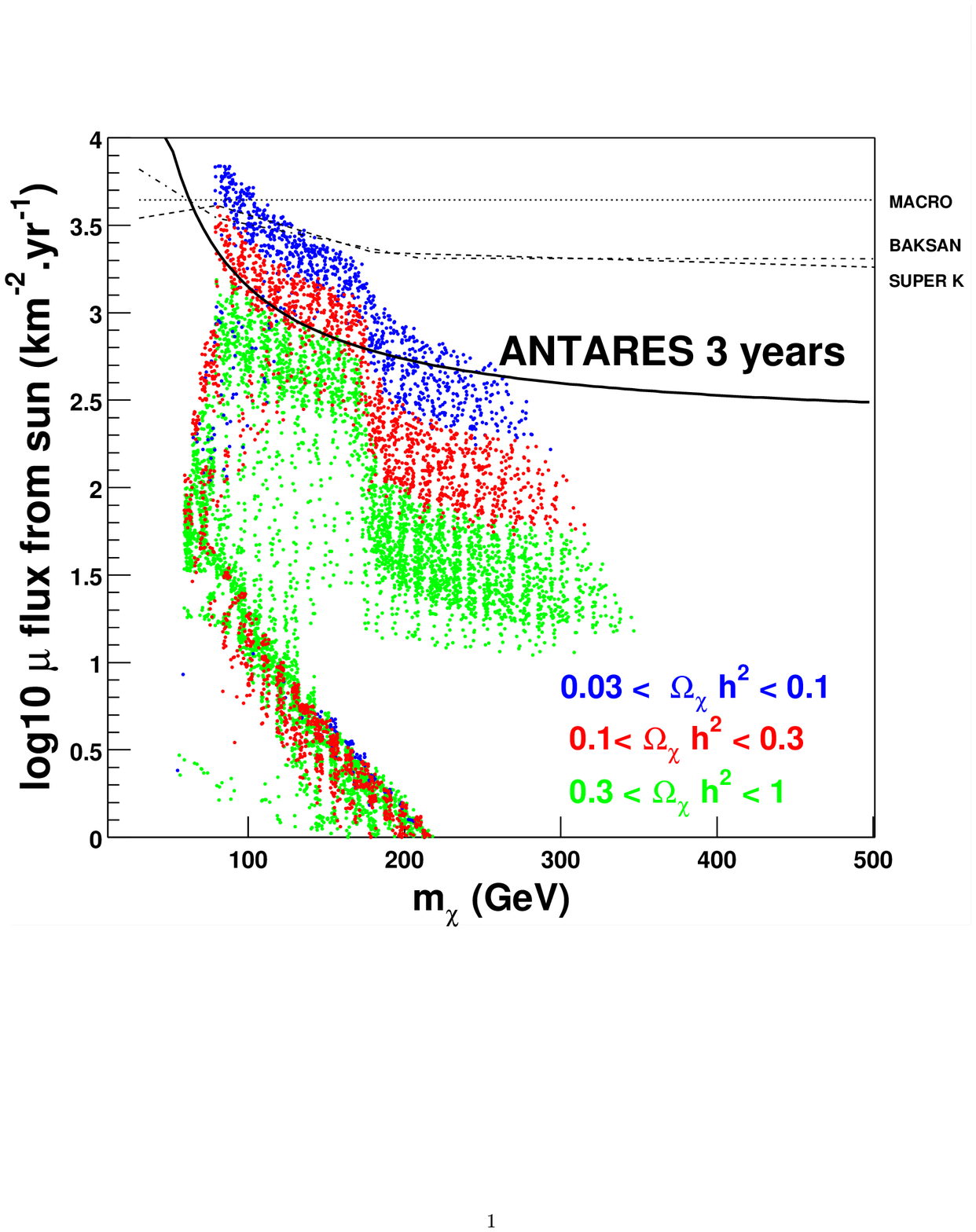}
\label{fig-dm}
\vspace{0.2cm}
\parbox{7.2cm}{\footnotesize Figure 2: ANTARES sensitivity (1 year of running)
to the neutrino-induced muon 
flux from astrophysical point sources, compared to upper limits from 
other experiments.}
\hspace{0.5cm}
\parbox{7.8cm}{\footnotesize Figure 3: ANTARES sensitivity to the muon flux from
neutralino annihilations in the centre of the Sun, compared to upper
limits from other experiments and mSUGRA predictions.}
\end{center}
\end{figure}  

\vspace{-0.5cm}
\section{Recent progress}

The first component of the final detector configuration, the main
electro-optical cable, was deployed in October 2001 and 
connected to the shore
station. In December 2002, the end of the cable was connected to the JB. 
Communication with the JB slow
control is being maintained since then.
Two prototype strings, an optical line with 5 storeys (Prototype Sector
Line, PSL) and the Mini Instrumentation Line (MIL -- with instruments 
to monitor environmental parameters), were deployed in
December 2002 and February 2003, respectively, and connected to the JB in 
an undersea
operation by the manned submarine Nautile in March 2003. Communication
with both lines was established immediately after connection. 
The systems were
found to be functional with the exception of two failures outlined below. 
A large quantity of data
was acquired and analysed to study the background rate and 
environmental parameters as functions of time. The lines have been recovered in
May (MIL) and July (PSL) 2003.

Two problems occured in the prototype tests. In the MIL, 
a water leak developed
in one of the electronic containers due to 
a faulty supplier specification for a connector; 
the design has been modified since then to exclude this
problem in the future. The second problem was that the clock
signal, sent from shore to the electronic modules to synchronise the
readout, reached the bottom string socket (BSS) but not the OMs. 
The clock failure was due to a damaged glass fibre in the cable between BSS and
first storey, caused by the supplier's use of an unsuitable material for 
the fibre coating.

Due to the absence of the clock signal, no data with timing information at
nanosecond precision could be taken. Nevertheless, the long-term operation of
the PSL over more than 3 months yielded a lot of information, both on the
functionality of the detector and the environmental conditions. In particular,
the rate of signals above threshold was monitored continuously for each OM. It
was found that the rates exhibit strong time variations, which are attributed
to bioluminescent organisms. A continuous rate, varying between about 
50 kHz and 250 kHz per OM, is accompanied by short light bursts. 
The fraction of time covered by the bursts varies from 
less than $1\%$ to more than $30\%$. Also monitored were
the position and tilt of the PSL storeys. It was found that they move almost
synchronously, i.e. the PSL behaves as a pseudo-rigid body in the water
current. Correlations of the background rates with the movement of the PSL and
hence with the sea currents have been observed. Detailed investigations of the
on-line filter requirements imposed by high rates and of relations between water
currents, string movements and bioluminescence are under way.

\vspace{-0.2cm}
\section{Summary and future plans}

With the installation of the main electro-optical cable and the junction box --
key elements for detector infrastructure -- 
the ANTARES project has entered the construction phase. Prototype detector
strings have been successfully deployed, connected to the JB,
operated and recovered, verifying the detector
design and functionality, yielding a vast amount of environmental
data and helping to identify and solve some problems. 
Failures, which occured in one connector
and in the transmission of the clock signal will be avoided in future by
implementing small design modifications. Intensive R\&D and 
environmental studies have been performed. The completion 
of the 12 string detector is scheduled for 2006.

%
%

\end{document}